\documentclass[12pt]{article}
\usepackage{latexsym}
\usepackage{amssymb}
 
\usepackage{mathrsfs}
\usepackage{amsmath}
\usepackage{epsf}

\def\pa{\partial} 
\def\half{\frac{1}{2}}

\begin{document}
\thispagestyle{empty}

\begin{center} 
  {\large \bf The role of dissipation in biasing the vacuum selection
in quantum field theory at finite temperature}\\[5ex]

F. Freire\,${}^{a}$\footnote{E-Mail:\ freire@lorentz.leidenuniv.nl},
N. D. Antunes\,${}^{b}$\footnote{E-Mail:\ n.d.antunes@sussex.ac.uk},
P. Salmi\,${}^{ab}$\footnote{E-Mail:\ salmi@lorentz.leidenuniv.nl}
and A. Ach\'ucarro\,${}^{a}$\footnote{E-Mail:\ achucar@lorentz.leidenuniv.nl}
\\[2ex]

   ${}^{a}$ Instituut-Lorentz, Universiteit Leiden,\\ 
   P.\,O. Box 9506, 2300 RA Leiden,\\ The Netherlands\\[1ex]
   ${}^{b}$ Centre for Theoretical Physics, University of Sussex,\\
   Falmer, Brighton BN1 9QJ, U.\,K.\\[4ex]

  {\small \bf Abstract}\\[2ex]
\begin{minipage}{14cm}
{\small
We study the symmetry breaking pattern of an $O(4)$ symmetric model
of scalar fields, with both charged and neutral fields, interacting
with a photon bath.
Nagasawa and Brandenberger argued that in favourable circumstances the
vacuum manifold would be reduced from $S^3$ to $S^1$.
Here it is shown that a selective condensation of the neutral fields,
that are not directly coupled to photons, can be achieved in the
presence of a minimal ``external'' dissipation, \textit{i.e.} not
related to interactions with a bath.
This should be relevant in the early universe or in heavy-ion
collisions where dissipation occurs due to expansion.\\[4ex]
     }
\end{minipage} 
\end{center} 

\newpage

\pagestyle{plain} 
\setcounter{page}{1} 


\section{Introduction and overview\label{1}}
In this paper we investigate the role that dissipation can play in
biasing va\-cuum selection after a symmetry breaking phase
transition. Our work was initially motivated by a mechanism suggested by
Nagasawa and Brandenberger \cite{Nagasawa:1999iv} to stabilise
non-topological classical solutions by out-of-equilibrium effects.
They studied an $O(4)$ model in which charged and neutral scalars
coupled differently to a thermalised photon bath.  Under the
assumption that only the charged fields receive thermal corrections
from the bath, these authors argued that the vacuum manifold would
collapse from $S^3$ to $S^1$ and therefore temporarily stabilise
non-topological strings.

The analysis in \cite{Nagasawa:1999iv} is focused on the stability of
embedded strings when immersed in a thermalised plasma. However, the
primordial question of whether the formation of the defect is dynamically
favoured has never been addressed. By looking at the requirements that
favour their formation we found a close link between ``external''
dissipation and vacuum selection that ranges beyond our initial aim. 
By external we mean a source of dissipation that is not related
to the interactions between the system and the heat bath.

The vacuum selection we discuss can take place in the early universe
and in heavy-ion collisions.
In the early universe, the most rele\-vant areas for applications are in 
the studies of preheating at the end of inflation \cite{Felder:2000hj} 
and defect formation in non-equilibrium cosmological phase transitions
\cite{Boyanovsky:1992vi}.

Current and future heavy-ion collision experiments also provide 
scena\-rios where dissipative vacuum selection might take place. 
A specific problem where this process might occur is the formation of 
disoriented chiral condensates \cite{Anselm:1989pk,Xu:1999aq}.  
Processes of similar nature might take place in the quark-gluon plasma
where modes with different thermalisation times coexist. However, 
our study, which is based on a symmetry breaking, does not
provide by itself a mechanism for this case. 

The mechanism we investigate is illustrated here for the same $O(4)$
scalar field theory in $3+1$ dimensions used in
\cite{Nagasawa:1999iv}. Some general assumptions are
required to specify the properties of the model but when presented out
of a particular context, at first sight, these might not appear natural. 
Therefore, we think it convenient to start with a brief outline of our
study.  In the context of our work we see the field theory as an
effective model for the soft long-wavelength modes of a system coupled to a
heat bath. At low temperatures the system has a symmetry broken phase 
and the symmetry is restored above some finite temperature $T_c$.

The following are key requirements concerning the coupling of the
scalar fields to the heat bath.  In the ordered phase, two of the scalar
fields, say for definiteness $\phi_{1,2}$, have decoupled from the
heat bath, while the remaining fields, say $\phi_{3,4}$, stay coupled.  
This situation is more natural than it seems at first sight. For
example it occurs when the fields $\phi_{1,2}$ are neutral and
$\phi_{3,4}$ are charged, both with respect to the same conserved
charge, and the heat bath consists predominantly of quanta of the
associated gauge field.

Furthermore, the coupling between $\phi_{3,4}$ and the bath is assumed
to be much stronger than the scalar self-coupling. Then, we expect
that by the time $\phi_{3,4}$ thermalise at the heat bath temperature
$T<T_c$, the $\phi_{1,2}$ fields are still
out of equilibrium. In other words, the relative strength of
the couplings implies that the relaxation times of $\phi_{3,4}$ are
much smaller than the ones of $\phi_{1,2}$. This is a condition that
can be realised in the early universe. The less favourable transition
is probably the most recent one, the chiral symmetry
breaking transition, because of the strong pion interactions.
Under these circumstances, when some fields are coupled and other
uncoupled to a heat bath, the question then arises: What is the
\textit{effective} vacuum manifold in this system?

When all the fields are coupled to the heat bath and dissipate
according to the fluctuation-dissipation relation the answer is well
known.  The vacuum manifold is the three-sphere $S^3$ covered by all
the equivalent scalar configurations that minimise the free energy.
The soft modes des\-cribed by the fields do not form a closed
system because their energy is being exchanged with the
heat bath, \textit{e.g.} via collisions and decay channels, giving
origin to dissipation. Let us assume, as it is normally the
case, that these terms are characterised by viscosity coefficients
$\eta_i$ associated with each scalar $\phi_i$.

The situation we study here differs in two ways. First, the neutral fields 
are not coupled directly to a heat bath. Second, and most importantly,
we consider 
these decoupled fields to have ``external'' sources of dissipation.
For example, in a cosmological context this type of dissipation is
naturally associated with the expansion of the universe. As a result
the decoupled $\phi_{1,2}$ fields stabilise in steady states that can
be characterised by an effective temperature $T_{\mathrm{eff}} < T$, 
independently on whether these fields
condense or not.

The most interesting effect occurs when the ``external'' dissipation
is much smaller than the dissipation in the coupled fields due to their
interaction with the bath. We would expect a small external dissipation
to have only a negligible effect on the evolution of the fields.
However, this is in general not the case. We have that at vanishing scalar
self-coupling the limit $\eta_{1,2} \to 0$, \textit{i.e.} the
``external'' sources of dissipation are ``switched off'', is singular. 
This has the effect of changing the symmetry breaking pattern
\textit{even} for small values of $\eta_{1,2}$.  
In this limit the
neutral fields are selected to condense, therefore effectively
reducing the vacuum manifold from $S^3$ to $S^1$ as originally
suggested by Nagasawa and Brandenberger \cite{Nagasawa:1999iv}. 

In reality, the uncoupled fields are still receiving energy from the
bath and the role of the external dissipation is simply to release
this energy at a comparable rate. This rate is much smaller than the
corresponding values for the coupled fields and has no appreciable
effect on them.

The origin of the vacuum manifold reduction in our analysis 
is identified without having to call for out-of-equilibrium effects as
in \cite{Nagasawa:1999iv}. 
We emphasise the important role played by the existence  of different
steady states for the various fields due to the ``external'' source of
dissipation.

The vacuum selection takes place above a small critical
dissipation which occurs when the neutral fields stabilise at a
``cold'' enough $T_{\mathrm{eff}}$. Our conclusions, therefore, is
that the reduction is more widely applicable than previously thought.

Our simulations are governed by phenomenological Langevin equations
describing the dynamical evolution of the fields. These equations have been
previously used in a relativistic context to study non-equilibrium
phenomena in cosmological phase transition
\cite{Gleiser:1994cf,Antunes:1997pm,Rivers:2004ir}. 
There are known limitations to the use of these equations which we
discuss in Sect. 4 and 5. They provide nevertheless an economic and
qualitative good description of the different processes involved
in the dynamic evolution of the fields where the coupling to the heat
bath is expressed by rapidly 
fluctuating fields and the dissipation effects are expressed by
viscosity terms.  In particular, this makes it easy to analyse the
effects of dissipation terms that are not related to interactions
with the heat bath.

The paper is organised as follows.  In Sect. 2, we discuss
the role of decoupling and dissipation in the vacuum biasing
mechanism.  A toy model illustrating the dissipation requirement in a
solvable system is presented in Sect. 3.
We illustrate in this simple case the effects of non thermal
dissipation terms that are not related to exchanges 
with a heat bath but originate for example from the expansion of the system. 
In Sect. 4 we study dissipative vacuum selection in the $O(4)$ symmetric
model in 3+1 dimensions. This section is divided in four subsections.  
We begin with a discussion on the use of phenomenological Langevin equations.
The results of our simulations are then presented in
Sect. 4.2 and in Sect. 4.3 we analyse the effect of varying the
parameters of the model. 
Finally, in Sect. 4.4 we discuss the conditions that guarantee the
vacuum selection to be in place when all fluctuating (and dissipative)
contributions are accounted for.
We end in Sect. 5 with a summary and a discussion of future work. 


\section{Decoupling and dissipation\label{2}}

We analyse the field dynamics in a system that undergoes a 
symmetry breaking transition and where different fields sectors in the
system reach distinct ``thermalised'' states with some heat bath.
The nature of the bath will be characterised below. Some fields
arrive at a standard thermalised state at the temperature of the bath
after a relatively short relaxation time.
The remaining fields stay out of equilibrium for a longer period,
which can still be small compared to observation times. 
The fields that take longer to either thermalise or reach
another type of equilibrium state are weakly or indirectly coupled to
a heat bath. We refer to them as \textit{decoupled} in a convenient
loose sense that will be made clear later.
In particular, we are interested in the situation where the 
decoupled fields condense following a finite temperature phase
transition. In order for this to happen these fields must  
lose most of their energy and for this reason we will follow
closely the role of dissipation.

In order to discuss a setting where this scenario can be realised we
use an $O(4)$ linear sigma model Lagrangean,
\begin{equation}
  \label{sigma-model}
  {\mathcal L} = \half\ \pa_\mu\phi_i\pa^\mu\phi_i\ - 
  \frac{\lambda}{4} \left(\phi_i\phi_i -
  v^2\right)^2\!,\ i=1,2,3,4,
\end{equation}
to describe the propagation and self-interactions of soft modes. 
For convenience, we consider $\phi_1$ and $\phi_2$ to be neutral
scalars and $\phi_3$ and $\phi_4$ to be the constituents of a charged
scalar $\phi^\pm = (\phi_3\pm i\phi_4)/\sqrt2$ with respect to a
$U(1)$ charge. For concreteness let it be $U(1)_{\mathrm{EM}}$.
Therefore, $\phi^\pm$ is coupled to a bath of photons while the
neutral scalars are not. In a more comprehensive ana\-lysis the effects
of the fluctuations from the hard modes of the scalar fields are also
to be taken into account.  We will discuss their effects in the next
sections.

By keeping the model simple we can aim at a better understanding
on how a small dissipation can play a role in selecting the
vacuum. This is the effect we wish to emphasise and alongside we lay
the conditions under which the vacuum manifold shows the selectiveness
that favours the formation of such embedded structures.  

Throughout this paper we work under the assumption that the coupling
between the charged scalars and the photon bath is much stronger than
the scalar self-coupling. 
Without this assumption leading to an effective sepa\-ration of scales no
non-trivial selection seems to take place.
With these conditions, we expect the charged scalars to
thermalise ``quickly''.  Their relaxation time sets the scale for what
we will refer as a quick thermalisation time.  In practice, at the
observation scale, the charged scalars can be said to remain in
equilibrium. The neutral scalars have of course a longer thermalisation 
time and are at least close to thermal equilibrium. 

One of our aims is to express quantitatively the distinction between 
the steady state reached by the neutral and the charged fields.
If the decoupled fields have no direct process to dissipate their
energy they reach a thermal equilibrium state at the temperature
$T$ of the photon bath. This therma\-lisation occurs because
the decoupled fields are not completely cut off from the photon bath 
due to the quartic scalar self-interaction. The rate at which the
neutral fields thermalise depends on the strength of
$\lambda$. Elsewise, if they dissipate due to the expansion of the
system as it cools, as in the early universe or heavy-ion collisions,
the steady state they reach is ``colder''.
This effect leads to their selective condensation and suggests the
use of an effective temperature $T_{\mathrm{eff}}$ as a way to
parametrise the distinct steady states. We will present a
detailed discussion of $T_{\mathrm{eff}}$ in the next sections.

Before ending this section we refer the reader to the further underlying view
that the Lagrangean \eqref{sigma-model} is better suited for the
broken phase.  Our model can be seen as a sector in a larger theory.
For instance above $T_c$ other mediating bosons that decouple at the
transition can be responsible for maintaining all the scalars in thermal
equilibrium. A known example that inspires this view can be found in
the electroweak transition. In this case the $W$- and $Z$-bosons
become massive below the transition while the photons remain massless
through the transition.

Renormalisation due to radiative and thermal corrections coming from
hard modes could also have been included in \eqref{sigma-model} but
this should not have a major effect in our phenomenological analysis.
More important are the fluctuation-dissipation effects coming from
the scalar hard modes.  They do indeed have some effect in the main
simple picture laid down in this section but they do not change
the outcome if the relatively magnitudes of the couplings remains as
assumed above as we show at the end of the next section.

Before we address the dynamical description of the model described by
\eqref{sigma-model} we look at a simple two particle system where
we can perform analytic calculations that highlight the main
features of the ``vacuum biasing''.

\section{Biasing in a two field system in zero dimensions\label{2b}}

In this section we illustrate how decoupling and dissipation combine
to bias the distribution of energy in a toy system.
We have two coupled one-dimen\-sional oscillators of unit mass and
restoring constants $m^2_0$ and $m^2_c$. Alternatively, they can be
interpreted as two interacting fields in zero spatial 
dimensions.  The system can be seen as a first approximation for an
effective model for the zero Fourier components of a field theory in
three spatial dimensions.  It is a simplified version of the $O(4)$
model introduced in the previous section, but now with only two fields.
Here one ``charged'' field $\phi_c$ is coupled to a heat bath, and the
other ``neutral'' field $\phi_0$ is not coupled.

We adopt a Langevin description, in analogy to what we will do in the
next section for the system described by \eqref{sigma-model}, where
the coupling to the bath is represented by a random rapidly varying
field. In a canonical form the equations of motion are,
 \begin{eqnarray}\label{field-eqn}
  \dot{\phi}_0&=&\pi_0 \nonumber \\
  \dot{\pi}_0&=&-\eta_0 \pi_0 - m_0^2 \phi^{}_0 -
              \frac{\partial V}{\partial \phi_0} \nonumber \\ 
  \dot{\phi}_c&=&\pi_c \nonumber \\
  \dot{\pi}_c&=&-\eta_c \pi_c - m_c^2 \phi^{}_c -
              \frac{\partial V}{\partial \phi_c} + \xi .
 \end{eqnarray}
The interaction between the fields is contained in the potential
$V(\phi_0,\phi_c)$  and the dissipation is expressed by the viscosity
terms involving the coefficients $\eta_0$ and $\eta_c$. The
interaction of the charged field, $\phi_c$, with the ``photon'' bath is
modeled by the random field $\xi$,
\begin{equation} \langle \xi(t) \rangle = 0,\,\,\,\,\,\,\,\,\,
\langle \xi(t) \xi(t') \rangle = \Omega \delta(t-t'),
\end{equation}
where $\Omega$ is the variance of the Gaussian white noise.
This ensemble of fields can be described by the probability
density $\rho(\phi_0, \pi_0, \phi_c, \pi_c,t)$. The evo\-lution 
of this distribution is governed by the Fokker-Planck type equation,
\begin{eqnarray}\label{fokker-plank}
 \frac{\partial \rho}{\partial t}&=& - \pi_0 
  \frac{\partial \rho}{\partial \phi_0}+
 \Big(m_0^2 \phi_0 + \frac{\partial { V}}{\partial \phi_0}\Big) 
 \frac{\partial \rho}{\partial \pi_0} 
 +\eta_0 \frac{\partial(\pi_0 \rho)}{\partial \pi_0}- \nonumber \\
 & &-\pi_c \frac{\partial \rho}{\partial \phi_c}+
 \Big(m_c^2 \phi_c + \frac{\partial { V}}{\partial \phi_c}\Big) 
 \frac{\partial \rho}{\partial \pi_c} 
 +\eta_c \frac{\partial(\pi_c \rho)}{\partial \pi_c}
 +\frac{\Omega}{2} \frac{\partial^2  \rho}{\partial \pi_c^2}\ .
\end{eqnarray}   
For any physical observable $A(\phi_0,\pi_0,\phi_c,\pi_c)$ we have
\begin{equation}
 \langle A \rangle(t) = \int d\phi_0 d\pi_0 d\phi_c d\pi_c \
 A(\phi_0,\pi_0,\phi_c,\pi_c)  \rho(\phi_0,\pi_0,\phi_c,\pi_c,t),
\end{equation}
and its time derivative can be calculated using equation
\eqref{fokker-plank} and integrating by parts.  The equation
of motion for $\langle A \rangle$ then reads
\begin{eqnarray}
 \frac{\partial\langle A \rangle}{\partial t}&=&\langle
  \pi_0 \frac{\partial A}{\partial \phi_0} \rangle -
 \langle\frac{\partial { V}}{\partial \phi_0} 
     \frac{\partial A}{\partial \pi_0} \rangle
 - m_0^2\,\langle \phi_0 \frac{\partial A}{\partial \pi_0} \rangle
 - \eta_0\,\langle \pi_0 \frac{\partial A}{\partial \pi_0} \rangle +
  \langle \pi_c \frac{\partial A}{\partial \phi_c} \rangle - \nonumber \\ 
 &-& \langle\frac{\partial { V}}{\partial \phi_c} 
     \frac{\partial A}{\partial \pi_c} \rangle
 - m_c^2\,\langle \phi_c \frac{\partial A}{\partial \pi_c} \rangle
 - \eta_c\,\langle \pi_c \frac{\partial A}{\partial \pi_c} \rangle
+\frac{\Omega}{2}\,\langle \frac{\partial^2 A}{\partial \pi_c^2} \rangle.
 \label{eq2}
\end{eqnarray} 
In equilibrium we have $\partial_t \langle A \rangle=0$ for all
physical observables $A$. 
One way of obtaining a full description of the system in equilibrium
is to obtain all the expectation values of all combinations of powers of
the four quantities $\phi_0$, $\pi_0$, $\phi_c$ and $\pi_c$. 
These are the $N$-point functions for (classical)-fields in
equili\-brium. It is easy to see that except for the simple case of quadratic
potentials (where we expect the system to be solvable), these equations mix 
correlations of powers of different degrees in the dynamical variables. 
This leads to an infinite number of linked equations similar to the
Dyson-Schwinger hierarchy. Here we will consider only the 
quadratic case and take the interaction potential to be
\begin{equation}\label{quadraticpotential}
 V(\phi_0,\phi_c)= \lambda \phi_0 \phi_c.
\end{equation}
This choice will enable us to carry an analytic treatment. For the
case with a quartic interaction $\lambda\phi_0^2\phi^2_c$ we have
checked that the main features of the results are the same. 

For the potential \eqref{quadraticpotential}, the second order
correlation functions with $i\neq j$ satisfy the following system of
10 equations, 
\begin{eqnarray}
\phi_i^2 &\rightarrow& \langle \phi_i \pi_i \rangle = 0\,,\ \ \
\phi_0 \phi_c \ \ \rightarrow\ \ \langle \phi_c \pi_0 \rangle +
\langle \phi_0 \pi_c \rangle = 0 \nonumber\\
\pi_i^2 &\rightarrow& \lambda \langle \pi_i \phi_j\rangle+m_i^2 \langle \phi_i
\pi_i\rangle +\eta_i\langle \pi_i^2\rangle - \delta_{i2} \Omega/2 = 0
\nonumber\\ 
\pi_0 \pi_c &\rightarrow& \lambda(\langle \phi_c \pi_c\rangle+\langle \phi_0
\pi_0\rangle)+m_0^2\langle \phi_0 \pi_c\rangle + m_c^2\langle \phi_c
\pi_0\rangle 
+ (\eta_0-\eta_c)\langle \pi_0 \pi_c\rangle = 0 \nonumber \\
\phi_i \pi_i &\rightarrow& \lambda \langle \phi_i \phi_j\rangle + m_i^2\langle
\phi_i^2\rangle+\eta_i\langle \phi_i \pi_i\rangle -\langle
\pi_i^2\rangle=0 \nonumber \\ 
\phi_i \pi_j &\rightarrow& \lambda \langle \phi_i^2\rangle +
m_j^2\langle \phi_i 
\phi_j\rangle+\eta_j\langle \phi_i \pi_j\rangle -\langle \pi_i
\pi_j\rangle=0\ ,
\end{eqnarray}
which is closed because it involves only expectation values of
degree two in the dyna\-mical variables. After some work we can solve it
obtaining the average for the momentum of the decoupled field,
\begin{equation}
\langle \pi_0^2\rangle = \frac{\lambda^2 \Omega}{2}
 \left[ \frac{\eta_0 \eta_c (m_c^2-m_0^2)^2}{(\eta_0 +\eta_c)}
       + \lambda^2 (\eta_0+\eta_c) +\eta_0 \eta_c 
        (\eta_c m_0^2 + \eta_0 m_c^2) \right]^{-1}. \nonumber \\
\end{equation}
In the simpler case when $ m_0=m_c=m$ this reduces to
\begin{equation}
\langle \pi_0^2 \rangle = \frac{\lambda^2 \Omega}{2 (\eta_0 +\eta_c)
 (\lambda^2 + m^2 \eta_0 \eta_c)}\ .
\end{equation}
The other momentum is related to this one by
\begin{equation}
\langle \pi_c^2 \rangle = \frac{\Omega}{2\eta_c} 
  - \frac{\eta_0}{\eta_c} \langle \pi_0^2 \rangle.
\end{equation}
Next we assume that the dissipation for the coupled field and
the amplitude of the noise obey the fluctuation-dissipation relation,
$\Omega=2\eta_c T$, where $T$ is the temperature of the bath.
We now define the effective stabilisation temperatures
\begin{eqnarray}\label{violation}
T^{}_\mathrm{eff} =  \langle \pi_0^2 \rangle &=& 
        \frac{\lambda^2 \eta_c T}{(\eta_0 +\eta_c) 
 (\lambda^2 + m^2 \eta_0 \eta_c)}\ , \nonumber\\
T^{\prime}_\mathrm{eff} = \langle \pi_c^2 \rangle &=&
T-\frac{\eta_0}{\eta_c}\langle \pi_0^2\rangle. 
\end{eqnarray}
for the neutral and charged field respectively. In the case of zero
dissipation coefficient for the neutral field,
$\eta_0 = 0$, not only these temperatures are the same but also
$T^{}_\mathrm{eff} = T^\prime_\mathrm{eff} = T$ for any finite non
vanishing values of $\lambda$. On the other hand, when
$\eta_0 > 0$ the effective temperatures of the
asymptotic states of each field are different which justifies using the
quantities defined by \eqref{violation} as convenient parameters
to express quantitatively the distinct steady states. We note that
similar definitions have been introduced in glassy systems
\cite{glass}, however the $T_\mathrm{eff}$ we use here are constants
as they refer to steady states after relaxation instead of long
lasting transient states. 

In this toy model the biasing mechanism corresponds to an unequal
distribution of the kinetic energy between the different types of fields.
The main features that characterise it are easy to identify
from equations \eqref{violation}. To  start with, note that if
$\eta_0\ll\eta_c$, $T^{\prime}_\mathrm{eff}$ is hardly 
affected with relation to its value at $\eta_0 = 0$, 
whereas $T^{}_\mathrm{eff}$ can be noticeably reduced,
provided $\lambda^2 \lesssim m^2 \eta_0 \eta_c$.
When these two conditions are simultaneously satisfied even a very
small value of the ratio $\eta = \eta_0/\eta_c$ can cause a large effect on
the asymptotic configuration of the uncharged fields $\phi_0$.
We have verified that simulations for a model with quartic
self-interactions indicate the existence of analogous relations.

The reason why a small value of $\eta$ is able to lead to a qualitatively
different behaviour is the existence of a singularity to which
we now turn our attention.  Consider the two sequences of limits,
$\lambda\to0$ followed by $\eta_0\to0$ and its reverse when $\eta_0$
is taken to zero first. In the former we have $T^{}_\mathrm{eff}\to0$
and $T^\prime_\mathrm{eff}\to T$, and in the latter both
$T^{}_\mathrm{eff}$ and $T^\prime_\mathrm{eff}$ approach $T$. Clearly,
the phenomena that we are studying corresponds to a regime along
the first sequence of limits where the two types of fields are weakly
coupled to each other but a dissipation in the neutral field prevails
even after the self-coupling is ``switched-off''. Therefore,
we are not only considering $\eta$ small but also $\lambda$ has to be
``small'' in the sense that
\begin{equation}
\frac{\lambda^2}{m^2 \eta_c^2} \lesssim \eta=\frac{\eta_0}{\eta_c} \ll
1. \label{biasing-condition}
\end{equation}
We might expect that the physical origin of the sizeable effect
for small $\eta$ follows from $\eta_0$ not being associated with
fluctuation-dissipation effects. But this by itself is not sufficient
to explain what we observe.  Equally important is the existence of at
least two scales in the problem. In order to clarify this point
let us consider the case when the neutral field also evolves under
the effect of fluctuations. To this end we modify the second equation
in \eqref{field-eqn} which now reads 
\begin{equation}
  \label{field-eqn-with-fluct}
    \dot{\pi}_0 = -\eta^{}_0 \pi_0 - m_0^2 \phi^{}_0 -
              \frac{\partial V}{\partial \phi_0} + \xi^{}_0,
\end{equation}
where the dissipation coefficient $\eta_0$ is now decomposed into two 
terms, $\eta^{}_0 = \eta^{\mathrm{ext}}_0 + \eta^{\mathrm{fl}}_0$. The
first term corresponds to an external source of dissipation while the
second one is related to fluctuating forces according to a
standard fluctuation-dissipation relation,
\begin{equation} \langle \xi^{\mathrm{fl}}_0(t) \rangle 
  = 0,\,\,\,\,\,\,\,\,\,
\langle \xi^{\mathrm{fl}}_0(t) \xi^{\mathrm{fl}}_0(t') \rangle 
  = \Omega_0 \delta(t-t'),
\end{equation}
where $\Omega_0 = 2\eta^{\mathrm{fl}}_0 T_0$, with $T_0$ the
temperature of the bath coupled to the neutral field. It is natural
in a first approximation to take $T_0 = T$ as we should not expect
a selective behaviour for the high energy modes. Moreover, this
situation corresponds to a case where biasing is less favourable.
The strength of the coupling of each field to the heat bath reflects
itself in the relative magnitudes of the dissipation coefficients. 
This matter will be made clearer in the full $O(4)$ model. 

A generalisation of the calculation leading to \eqref{violation}
including the noise for the neutral fields and with $T_0 = T$ gives
\begin{equation}
  \label{eq:new-temperature-diff}
  \frac{T^\prime_{\mathrm{eff}} - T^{}_{\mathrm{eff}}}{T} = 
   \frac{\eta^{\mathrm{ext}}_0}{\eta^{}_0}
  f(m^2\eta^{}_c\eta^{}_0/\lambda^2),
\end{equation}
with $f(x)=x/(1+x)$, from which we see that the effective temperatures
of the steady states can still differ substantially if in addition to 
\eqref{biasing-condition} we have
$\eta^{\mathrm{ext}}_0 \sim \eta^{}_0$. These combine to guarantee
that the argument of $f$ is large therefore assuring that $f\sim 1$. 
Under these conditions
$T^\prime_{\mathrm{eff}} - T^{}_{\mathrm{eff}} \sim T$.

Note that an external dissipation should for consistency affect both
fields. The inclusion of such non-thermal dissipation in the equations of
the charged fields adds a new term in the right-hand side of
\eqref{eq:new-temperature-diff}, analogous to the present one, where 
for instance the prefactor is now $\eta^{\mathrm{ext}}_c/\eta_c$
instead of $\eta^{\mathrm{ext}}_0/\eta_0$.
As $\eta^{\mathrm{ext}}_c = \eta^{\mathrm{ext}}_0$ this leads 
to a small, ${\cal{O}}(\eta_0/\eta_c)$, reduction in the
thermalisation temperature of the charged fields and no 
qualitative change.

Therefore, it is justifiable, under the 
requirements at the end of the last paragraph together with
\eqref{eq:new-temperature-diff}, to neglect the external dissipation
contribution to the charged field as
$\eta_c^{\mathrm{ext}} \sim \eta_0^{\mathrm{ext}} \ll \eta_c^{}$. 
It is the combination between the presence of an 
external source of dissipation and the existence of two scales,
say $\eta_c^{\mathrm{fl}}$ and $\eta_0^{\mathrm{fl}}$, that is behind
the physical origin of the effect we are analysing. 
The effect is only noticeable when the external dissipation is at
least comparable to the smaller scale. 
In the next section we carry out a more explicit discussion on the
role of these scales. 

We conclude the study of this toy model with a couple of
graphical illustrations of the distinct steady states and their
associated relaxation times.  For convenience we go back to the starting
study case, $\xi_0 = 0$ and $\eta^{}_0=\eta_0^{\mathrm{ext}}$, with
the system governed by equations \eqref{field-eqn}.

\begin{figure}
\centerline{\epsfxsize=3.8in\epsfbox{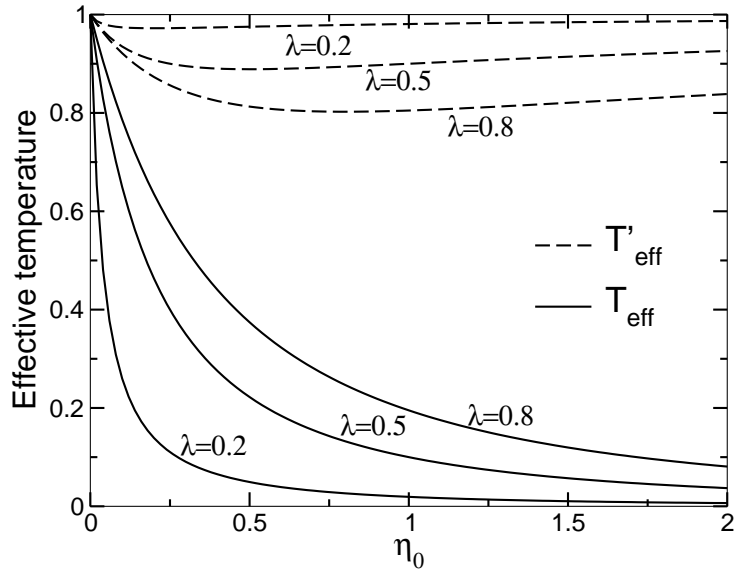}}   
\vskip-.3cm
\caption{\label{inter1} The effective stabilisation temperatures for
  each of the fields as a function of the viscosity coefficient
  $\eta_0$. The curves shown are for three values of $\lambda$ and 
  $m = T = \eta_c = 1$.}
\vskip-.3cm
\end{figure}
Equations \eqref{violation} are represented graphically in Figure
1 for three different values of $\lambda$.  Clearly, as $\eta_0$
increases from zero, $T_\mathrm{eff}$ decreases from the starting
value $T$. The drop in temperature is more pronounced when the self
coupling is smaller.  $T_\mathrm{eff}$ approaches
negligible values for $\eta_0 \gtrsim \eta_c$ and sufficiently small
$\lambda \lesssim 0.2$. This simply reflects that when the indirect 
interaction of $\phi_0$ with the heat bath is weaker this field
dissipates its energy more efficiently. 
On the other hand, the field $\phi_c$ because of its
direct coupling to the heat bath remains ``hot'', \textit{i.e.}
$T^\prime_\mathrm{eff}$ stays close to $T$, but the larger
$\lambda$ is, the more it deviates from $T$. 
This deviation is a natural consequence 
of the direct dissipative viscosity term for
$\phi_0$ which is not balanced by any fluctuating effects. 
Although not shown in Figure 1 we have from \eqref{violation} that when
$\lambda \to \infty$ the effective temperatures become identical for
any value of $\eta_0$.

\begin{figure}
\centerline{\epsfxsize=3.8in\epsfbox{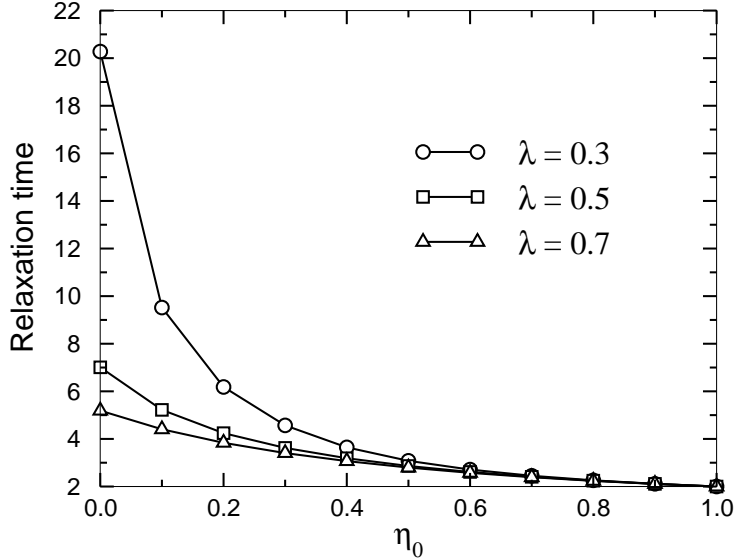}}   
\vskip-.3cm
\caption{\label{inter2} The relaxation times for $\phi_0$ as a
  function of the viscosity coefficient $\eta_0$. The curves are for
  three values of $\lambda$ and $m = T = \eta_c = 1$.}
\vskip-.3cm
\end{figure}

The relaxation times for the decoupled fields for various values of
the self-coupling are shown in Figure 2.  
The results were obtained by determining the eigenvalues of the homogeneous
version of equations \eqref{field-eqn} (\textit{i.e.} 
ignoring the noise term). The several relaxation time scales for the
system are proportional to the inverse of the real part of these
eigenvalues. In Figure 2 we show for each value of $\eta_0$ the
largest of these time scales, which we interpret as the equilibration
time for the uncoupled fields. We also checked that for all values of
$\eta_0$ and $\lambda$ there is one eigenvalue leading to a relaxation
time close to $2.0$, corresponding to the charged fields. When the two
values of the dissipation are the same, for $\eta_0=\eta_c=1$, the two
time scales coincide as expected.
The rapid slowing down in the rate at which $\phi_0$ stabilises as
$\lambda$ decreases shows a dependence closer to $\lambda^{-2}$ as we
would expect. This naive expectation derives from the effective noise
term ${\cal O}(\lambda)$ that the interacting potential induces in the
equation for $\phi_0$ in the equations of motion \eqref{field-eqn}. 

In summary, by looking at a two field toy model we observe an unevenness
in the way the kinetic energy is distributed between the ``charged'' and
the ``neutral'' field.  The conditions for this biasing are possible
because of the presence of an external dissipation term and different
strengths for the coupling of the fields to the heat bath. 
When the field with the weakest coupling to the
bath is decoupled we recognise that this counterintuitive behaviour is
due to a singularity in the limit of vanishing external dissipation.
In a realistic situation the weakest coupling to the bath should not
be neglected.  The singular gives place to a two scale regime and the
biasing is expect to occur when $\eta^{\mathrm{ext}}$ extracts energy
at least at a rate comparable with the input from the bath coming from
the weaker coupling. 
In the next section we will see how a similar effect contributes to a
vacuum selection following a symmetry breaking phase transition. 


\section{Vacuum biasing in the $O(4)$ model\label{3}}


\subsection{The use of the Langevin approach}
We now return to the model described by the Lagrangean \eqref{sigma-model} 
and we will study its dynamical evolution under the Langevin approach
already used in Sect. 3.  In order to simulate the dynamics of the
fields we use phenomenological Langevin equations 
\begin{equation}
  \label{langevin}
  \Big[(\pa_t^2-\nabla^2) - \mu^2 + \lambda\sum_{k=1}^4\phi_k^2 +
  \eta_i\pa_t\Big]\phi_i  = \xi_i,
\end{equation}
with $\mu^2 = \lambda v^2$, $v$ being the $T=0$ vacuum expectation value,
and where $\eta_i$ and $\xi_i$ are respectively the viscosity coefficients
and the Gaussian noises. For the fields that couple to the photon bath
and that are assumed to thermalise at its temperature $T$, we have
\begin{equation}
  \label{noise}
  \langle \xi_i(\vec x,t) \xi_j(\vec x',t') \rangle =
  \Omega_i\delta_{ij}\delta^{(3)}(\vec x-\vec x')\delta(t-t').
\end{equation}
where $\beta\Omega_i = 2\eta_i$, with $\beta=1/T$, according to the
fluctuation-dissipation theorem.  Below $T_c$ this relation is used
only for $i=3,4$, while in the disordered phase it is assumed for
all the fields. 

In the ordered phase we consider the neutral fields to be decoupled from
the bath, $\xi_i = 0$, but let $\eta_i \neq 0$.  The non vanishing value 
of these coefficients are due to an ``external'' source of dissipation.
An expansion of the system, as in the early universe, is a possible
origin for this type of dissipative term. In fact, this has been the
main motivation behind this study.
The toy model in the previous section showed us that even a small
external dissipation could give rise to non negligible effects. In
cosmology we expect a similar situation (except at very early times)
in the sense that the dissipation due to the expansion of the universe
is much smaller than the one associated to the charged fields which is
due to the interactions with the ``photon'' bath. 

It is in place to say something on the advantages and limitations of
using the Langevin equations \eqref{langevin} at this stage.  These
equations describe the classical out-of-equilibrium evolution of a
system of coupled fields. Therefore, at most, it provides effective
equations for the long-wavelength modes of quantum fields with large
occupation number. For this reason it is often used as a
phenomenological set of equations to study close-to-equilibrium effects
near phase transitions \cite{Goldenfeld:1992qy}. The use of the
relativistic form of these type of equations has been motivated by
cosmological applications \cite{Gleiser:1994cf,Antunes:1997pm}.

The relative simplicity of these equations is their main asset.  We
take the noise to be white and the dissipative kernel local both in
space and 
time. Nevertheless, they provide a good starting set of equations to
look for qualitative answers to questions in close-to-equilibrium dynamics
provided one keeps a careful perspective of its shortcomings. This is
the approach we take here.  

The shortcomings of equations \eqref{langevin}
are best understood when a more systematic derivation of the effective 
out-of-equilibrium dynamics from first principles is carried out
\cite{Gleiser:1993ea,Greiner:1996dx,Csernai:1997ye}. The
phenomenological approach to the study of hot non-Abelian plasmas
provides also useful insights \cite{Bodeker:1998hm,Litim:1999ca}.       
There are, for instance, the questions on whether the origin of the
noise is external or internal \cite{Csernai:1997ye},
or that the Langevin approach is only
reliable near thermal equilibrium when a quasi-particle type of
approximation can be justified. Let us say something briefly on the
second issue.

One of the safest ways of understanding the regimes where a Langevin 
\mbox{description} applies in quantum field theory is to start from the
Kadanoff-Baym equations \cite{Danielewicz:1982kk} and work out under
which conditions these equations can be interpreted as the result of
Langevin processes \cite{Greiner:1998vd}.  The high temperature limit is
a well established condition and is naturally implicit in
\eqref{langevin} as this is an effective equation for long-wavelength modes,
$k\ll T$. The use of classical equations is also justifiable by the
more recent investigations on the reliability of the classical field theory
limit to the dynamics of quantum fields out of equilibrium
\cite{Aarts:2000mg,Arrizabalaga:2004iw,Rivers:2004ir} at high $T$ or
near $T_c$.


\subsection{The simulations}
We use a discretised version of \eqref{langevin} in three dimensional 
square lattices with $50^3$ to simulate the evolution of the $O(4)$
model \eqref{sigma-model}. A leap-frog algorithm with time step 
$\delta_t = 0.05$ is used. Larger lattices of $100^3$ have been used
to verify the stability of our results. A Gaussian random number
generator is used for the rapidly changing fluctuations.
All quantities are measured in units of the $T = 0$
vacuum expectation $v$.  The dimensionless quantities are identified
with a tilde.  For example, $\widetilde v =1$ and $\widetilde\beta = v/T$.
When choosing the lattice spacing we need to ensure that the modes
with wavelength longer than $\sim T^{-1}$ are not cut off.
In our runnings merely for reference we took physical scales from the
chiral symmetry breaking effective mean field model.
By using a lattice spacing $\widetilde{\Delta x} = 0.25$ for
$v = 93$ MeV we can work up to temperatures of approximately $T\sim350$ MeV.

We run our simulations for successive temperatures of the heat bath
determined by $\Omega_+ = 2 \eta_+ T$, the amplitude for the noise of
the charged scalars.  Starting from a temperature $T > T_c$ we bring
down the temperature across $T_c$. While in the disordered phase all
the fields are taken to be in contact with the heat bath.  It is only 
for $T < T_c$, when the vacuum expectation value starts to increase,
that the neutral fields are decoupled.

In Figure 3, we plot the order parameters for the condensates of neutral
and charged scalars against the inverse temperature $\beta$. These
are, respectively, 
$\langle|\phi_1\phi_2|\rangle$ and $\langle|\phi_3\phi_4|\rangle$, where
$\langle|\phi_i\phi_j|\rangle = \sqrt{\sum_{n=i,j}(\frac{1}{V}\int_V
  \phi_n(x))^2}$\ are ave\-rages over the entire lattice.
As the details of the decoupling of the neutral fields are not known
we show the curves for three 
different types of decoupling. The label of instantaneous, exponential
and power law refer to the way the variance $\Omega_0 = \Omega_0(T)$ 
approaches zero starting from its value at the disordered phase as $T$
continues to decrease below $T_c$. For the runnings in Figure 3, the
values of all the viscosity coefficients $\eta_i$ are kept the same 
for all the scalars.
The different decouplings do not have an effect on the final values of
the order parameters which always favour a non vanishing value for
$\langle|\phi_1\phi_2|\rangle$.
This shows a bias for the condensation of the neutral fields
resulting from the relative large value of the ratio between the
viscosity coefficient of the neutral by the charged scalar.
Here we used $\eta_0/\eta_+=1$ to emphasise the case when the condensation
of the neutral sector is strongly favoured. 

\begin{figure}[]
\centerline{\epsfxsize=3.5in\epsfbox{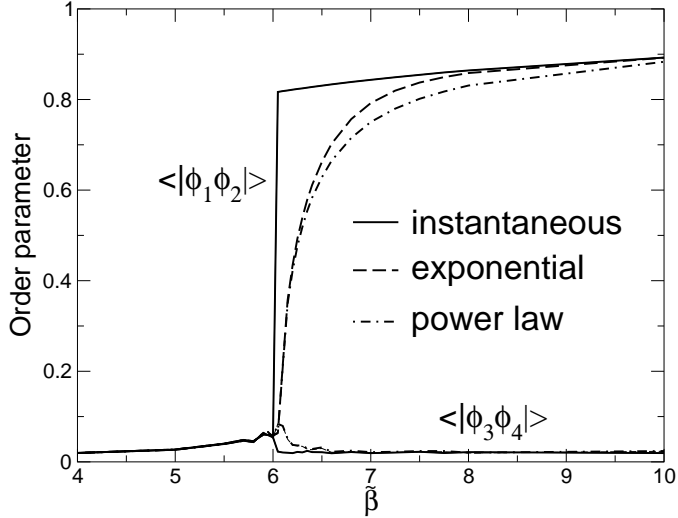}}
\caption{\label{inter3}Diagram showing the variation in the order
parameter as the system is cooled across $T_c$.  We use $\lambda = 1$ 
and a $50^3$ box. We show curves for three different decouplings:
instantaneous, exponential and  power law.}
\vskip-.3cm
\end{figure}

For the large value of $\eta_0$ used in the simulations for Figure 3
the neutral and the charged scalars have an equally effective channel
to cool as the temperature decreases.  However, the fluctuations
coming from the interaction with the heat bath have the effect of
slowing down the dissipation of the charged fields.  This will favour
the neutral fields to roll down more effectively to the bottom of the
potential and condense.

The neutral fields are not blind to the photon bath due to the scalar
self-coupling.  They dissipate through the viscosity term but they gain
energy via the scalar self-coupling.
However, as long as the effects of fluctuations hitting the neutral
modes is small and any increment of energy can quickly be dissipated,
which occurs when $\lambda$ is not too large and $\eta_0$ not
too small compared to $\eta_+$, the neutral scalars continue to
monopolise the vacuum.  As we discuss next the situation might change as
$\eta_0$ decreases.

The curves in Figure 4, where we set $\eta_0 = 0$ but all other
parameters are kept the same as in the runnings used for Figure 3,
show that in this case the charged scalars condense.  
This suggests that as $\eta_0$ is decreased below some critical
value the charged fields condense instead of the neutral ones. This
results in a ``superconducting'' background where the photons are
massive which clearly violates our working assumption of having a 
thermalised photon bath.  Therefore, we conclude that the condensation
of the charged scalars is an artifact of our simulations in this
region of the parameter space of the model.

\begin{figure}[ht]
\centerline{\epsfxsize=3.5in\epsfbox{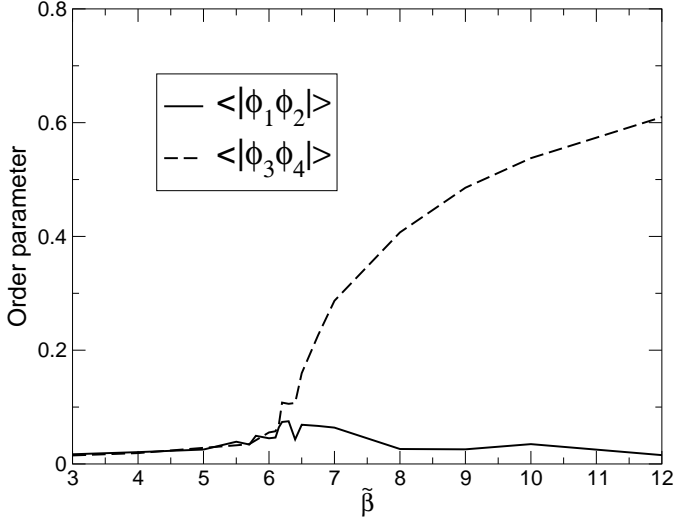}\hspace{.0\hsize}}
\vskip-.3cm
\caption{\label{inter4}Diagram showing the vacuum selection favouring
  the charged scalars to condense when $\eta_0=0$ in the broken phase.}
\end{figure}

From the simulations we have discussed, we anticipate the 
existence of a critical value $\eta_\mathrm{cr}$ for $\eta=\eta_0/\eta_+$ 
in the interval $0 < \eta_\mathrm{cr} < 1$. A precise
determination of $\eta_\mathrm{cr}$ is numerically delicate and at
this phenomenological stage of our study it does not justify the
dedicated effort it requires.
For certain this critical value indicates the end of the validity of the
context of our working conditions.  This, we expect, indicates
a qualitative change on the nature of the condensation.

The region of critical $\eta$ is characterised by competing domains
of neutral and charged scalars, which might present analogies to
disoriented chiral condensates \cite{Anselm:1989pk}.  In this region
we have estimated the value of $\eta_\mathrm{cr}$ for varying
self-coupling and present them in Table 1.  
Although the dependence with $\lambda$ is difficult to establish,
our results give an indication that as $\lambda$ is increased by
approximately one order of magnitude, $\lambda = 4 \to 20$, the
critical value of $\eta$ also increases by a similar magnitude,
$\eta_\mathrm{cr} \sim 10^{-5} \to 10^{-4}$. 
This is the behaviour we would naively expect from the fact
that the scalar self-coupling is also a measure of the extent to 
which the decoupled fields have indirect contact with the photon bath.
Finally, we remark that the results in the Table 1 for 
$\widetilde\beta = 9$ might already fall outside the region where a
classical description based on a Langevin approach is not reliable
\cite{Rivers:2004ir} as the system is no longer close to $T_c$. 

   \begin{table}[]
   \caption{The table shows the values of $\langle|\phi_1\phi_2|\rangle$
                       and $\langle|\phi_3\phi_4|\rangle$. The three
                       values of $\widetilde\beta$ correspond respectively
                       to a temperature above, near and below $T_c$.}
\begin{center}
\begin{tabular}{l|c|c|c|c|c|c|}
$\lambda=4$&\multicolumn{2}{c|}{$\widetilde\beta=4$}&\multicolumn{2}{c|}  
{$\widetilde\beta=6$}&\multicolumn{2}{c|}{$\widetilde\beta=9$}\\
       \hline\hline
$\widetilde\eta_0=1\cdot10^{-4}$ & .011 & .011 & .113 & .046 & .607 & .047\\
       \hline
$\widetilde\eta_0=5\cdot10^{-5}$ & .011& .011& .072 & .048 & .586 & .123\\
       \hline
$\widetilde\eta_0=2\cdot10^{-5}$ & .011& .011& .057 & .046 & .552 & .225\\
       \hline\hline\\[-.3cm]
$\lambda=12$&\multicolumn{6}{l}{}\\
       \hline\hline
$\widetilde\eta_0=2\cdot10^{-4}$ & .008 & .008 & .198 & .089 & .621 & .126\\
       \hline
$\widetilde\eta_0=1\cdot10^{-4}$ & .008& .008& .135 & .157 & .604 & .185\\
       \hline
$\widetilde\eta_0=5\cdot10^{-5}$ & .008& .008& .108 & .165 & .623 & .111\\
       \hline\hline\\[-.3cm]
$\lambda=20$&\multicolumn{6}{l}{}\\
       \hline\hline
$\widetilde\eta_0=5\cdot10^{-4}$ & .007 & .007 & .292 & .071 & .649 & .098\\
       \hline
$\widetilde\eta_0=2\cdot10^{-4}$ & .007& .007& .133 & .250 & .639 & .140\\
       \hline
$\widetilde\eta_0=5\cdot10^{-5}$ & .007& .007& .074 & .278 & .650 & .078\\
       \hline
\end{tabular}
    \end{center}
   \end{table}


\subsection{Interplay between parameters\label{3b}}

It is useful to analyse the interplay between the various
parameters in the model. To this end, we look at the kinetic energy
of each set of fields, the coupled and the decoupled. Parallels
with the toy model studied in the previous section will also be easier
to draw. With this
study we aim at learning how thermalisation is affected by the ratio
of the viscosity coefficients $\eta = \eta_0/\eta_+$ and the scalar
coupling $\lambda$.

In Figure 5 the time evolution of the kinetic energies for both types
of scalar fields are plotted.  The parameters $\widetilde\eta_0 = 0.005$,
$\widetilde\eta_+ = 1$ and $\widetilde\beta = 6.5$, all in units
of the $T = 0$ vacuum expectation value $v$ in a $100^3$ box, are the
same for all the curves, whereas we use three different values for
$\lambda$.

The equilibrium curve
corresponds to the coupled fields which thermalise quickly in the time
scales displayed in our plot. Of course ``quickly'' here 
means that the relaxation times for the decoupled
fields are clearly longer than for the coupled ones. This occurs
for all the parameter values shown. We verified that the
temperature for the equilibrium curve is to a very good approximation
consistent with the equipartition relation and therefore independent
of $\lambda$. The external dissipation
does not give origin to noticeable deviations from equipartition, as in
the toy model, because of the present large number of degree of freedom.

\begin{figure}[ht]
\centerline{\epsfxsize=3.8in\epsfbox{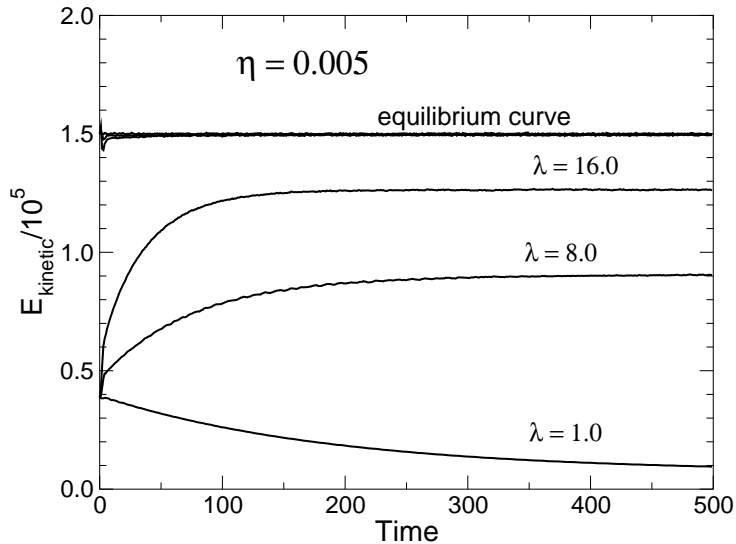}\hspace{.08\hsize}}
\vskip-.3cm
\caption{\label{inter}Diagram showing the kinetic energy of the scalar
fields for a single value of $\eta=\eta_0/\eta_c$ and three different 
values of $\lambda$. For small values of $\lambda$ the neutral scalars
approaches a steady state at a very slow rate.}
\vskip.3cm
\end{figure}

\begin{figure}[ht]
\centerline{\epsfxsize=3.8in\epsfbox{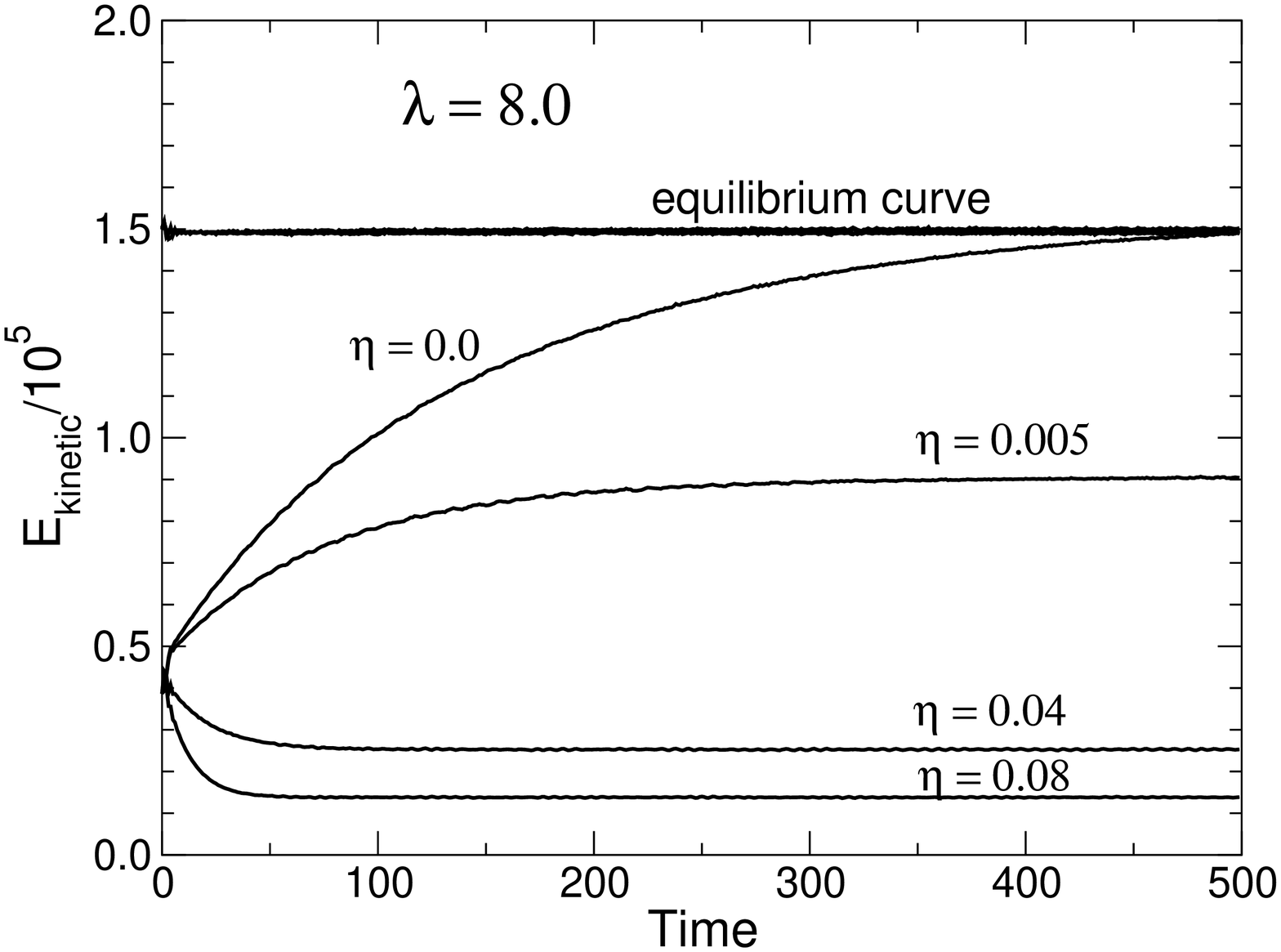}\hspace{.08\hsize}}
\vskip-.3cm
\caption{\label{inter}Diagram showing the kinetic energy of the scalar
fields for a single value of $\lambda$ and four different values of
$\eta=\eta_0/\eta_c$. The slowest relaxation time for the neutral
fields occurs in the $\eta=0$ limit  of no external dissipation.}
\end{figure}

The most interesting feature of Figure 5 is the $\lambda$ dependence
of the asymptotic values for the kinetic energy of the decoupled fields.
As in \eqref{violation} we can interpret these quantities as
\textit{effective equilibration temperatures} $T_\mathrm{eff}$.
We observe that the larger $\lambda$ is, the faster the decoupled
fields approach a steady state. This is to be expected 
as  the decoupled fields interact indirectly with the photon bath via
the quartic scalar coupling.  It explains not only the
shorter relaxation times for larger values of $\lambda$ but also the higher
$T_\mathrm{eff}$ values which get closer to the temperature of the bath.

We verified that the value of the asymptotic kinetic energies is
independent of the initial conditions. In Figure 5 the starting kinetic 
energy of the decoupled fields is a third of the equipartition value.
It is therefore safe to conclude that for weak scalar couplings the
equilibrium state should be quite distinct from the equipartition
thermalised state.

In Figure 6 we complement the curves shown in Figure 5 by keeping
now the same scalar coupling for all the curves, here we use
$\lambda = 8$, and vary $\eta_0$, or equivalently $\eta$ as we again
set $\eta_+ = 1$. We observe that the larger the viscosity coefficient 
$\eta_0$ the faster the decoupled fields equilibrate.  On the other
hand, as $\eta$ increases $T_\mathrm{eff}$ shifts away from $T$. The
asymptotic steady state approaches the 
equilibrium curve only in the opposite limit, \textit{i.e.} when the
external dissipative channel is ``switch off''.
In this case, the kinetic energy does eventually reach the value
expected by equipartition but at a clearly slow rate set by the
magnitude of $\lambda$ via the fluctuations mediated by the scalar
coupling.


\subsection{Coupling the decoupled fields\label{3c}}

We end this section with a discussion on the effects of introducing
fluctuation terms to the evolution equations for the decoupled fields
in analogy to what we have done in \eqref{field-eqn-with-fluct} for
the toy model.  
The Langevin equations already include indirect interactions of the
decoupled fields with the bath.  However, one may still wonder if
results might change when we include a more direct source of
fluctuations. 
The most natural reason for introducing additional
fluctuating forces is the interaction between the soft modes described
by the Langevin equations and the associated scalar hard modes
which should be present for both types of fields.

The relative strength of the different interactions is central here.
Let $g$ denote the gauge coupling.  Then our working assumption
that the scalar self coupling is weak compared to the coupling to the
photon bath means $g\gg\lambda$. In a simple perturbative estimate
this implies that $\eta_+\gg\eta_0^{\mathrm{fl}}$. Qualitatively this
relation is expected to hold at very high temperatures but more
reliable non perturbative statements require more dedicated
simulations.  

In order for the external dissipation to have observable effects it
can not be negligible with relation to both thermal dissipation
coefficients.  In our simulations this is trivially the case as
$\eta_+\gg\eta_0^{\mathrm{ext}} = \eta_0^{\mathrm{}}$, but 
$\eta_0^{\mathrm{ext}} > \eta_0^{\mathrm{fl}}=0$. The first of these
two inequalities justifies neglecting non thermal terms in the
equations for the coupled scalars.  In general, the biasing should
occur as long as
$\eta_0^{\mathrm{ext}} \gtrsim \eta_0^{\mathrm{fl}} \neq 0$.
Let us see how our analysis supports this view.

Looking back at equation \eqref{eq:new-temperature-diff} we see that
in the toy model the neutral field dissipation must be external for
the effective temperatures of the asymptotic states to be different.
A similar result applies to the $O(4)$ model where no vacuum selection
occurs unless there is an external dissipation.  From Figu\-re 6 we
recognise that it is necessary for the dissipation coefficient
$\eta_0$ to have a non thermal external component.
Clearly if $\eta_0$ had a purely thermal origin the kinetic
energy of the neutral field would always asymptotically approach the
equilibrium curve. This is expressed in the $\eta=0$ curve
in Figure 6 as thermal dissipative effects are implicitly already
present due to the scalar self-coupling.  Adding an explicit coupling
to a heat bath would only have the effect of decreasing the relaxation
time.


\section{Summary and discussion\label{4}}


In this paper we discussed the conditions that cause a selective
condensation of neutral fields in an $O(4)$ symmetric scalar field
sector of a field theory in three dimensions when this sector also
includes charged fields.
In particular, we studied a regime in the vicinity of a
symmetry breaking 
transition where the interactions with a thermalised photon bath can
be simulated by a phenomenological Langevin approach.

We show that under quite general conditions a non thermal dissipation
in the evolution of the neutral fields effectively reduces the vacuum
manifold of a system described by an $O(4)$ scalar model from $S^3$ to
$S^1$, above a small dissipation threshold. From this analysis we
identify the conditions that favour the 
formation and stabilisation of embedded defects as first argued by
Nagasawa and Brandenberger \cite{Nagasawa:1999iv}.

The vacuum manifold reduction is due to the existence of different
asymptotic steady states for the two types of scalar fields
considered. This effect is caused by an ``external'' source of
dissipation, in the sense that it does not arise from fluctuations
resulting from interactions with the photon bath. 
The neutral field steady state is characterised by an effective
temperature $T_{\mathrm{eff}} < T$, ``colder'' than the photon bath,
and this is what leads to their selective condensation.

The remarkable feature is that even a small amount of ``external''
dissipation can be sufficient to cause qualitatively distinct effects,
such as the vacuum selection. Small here is in relation to the
dominant dissipation terms in the charged scalar field sector, which
is related to the interaction with the photon bath by the
fluctuation-dissipation theorem. This is counterin\-tuitive. In
principle, one would be inclined to neglect the possible effects of
such a small amount of dissipation. For instance, the asymptotic
state of the charged field sector is hardly affected by the
``external'' dissipation.  What changes things is the existence of
a small indirect thermal dissipation. Then the
external dissipation needs at least to be of the order of the
much smaller thermal dissipation coming from the indirect coupling of
the neutral scalars with the heat bath.

We can naturally generalise the system to a more realistic one.  First,
the external dissipation is considered in both scalar field sectors.
However, for the charged field sector this leads only to 
negligible corrections. Second, the neutral scalars are coupled
directly to the heat bath although with a much weaker coupling than the
charged scalars so that the resulting thermal dissipation for the
former does not dominate over the non thermal external dissipation.
Under these general conditions our results on vacuum selection are not
qualitatively changed. 

Finally, we note that possible corrections to the scalar fields potential 
coming from the interactions with the gauge bosons should not play an
important role for our analysis. We know from the work of Nagasawa and
Brandenberger \cite{Nagasawa:1999iv} that the asymmetry created by the
decoupling from the neutral fields from the photon bath biases the
effective potential in a way that stabilises non topological defects
when immersed in a photon plasma. Moreover, equilibrium thermal
corrections to the potential tend to reduce the instability of these
embedded configurations 
\cite{Holman:1992rv}. Therefore, at a perturbative level we do not
expect corrections to counteract the vacuum selection we analyse here.
A less investigated difficulty, but potentially an important one,
is the contribution from very soft photons.  Because of infra-red
divergences reliable corrections similar to those in
\cite{Gleiser:1993ea,Greiner:1996dx} are not to our knowledge
currently available. More dedicated simulations including the full
dynamics of both the scalars and the gauge bosons are necessary
to clarify this problem.

The type of scenario we describe here should be relevant for
applications in the early universe where the expansion of the universe
provides a non thermal source of dissipation.  It would also be
interesting to investigate if similar effects might take place
in the quark-gluon plasma where different steady or intermediate states
might coexist as in the bottom-up thermalisation in heavy-ion collision
proposed in \cite{Baier:2000sb} driven by soft gauge bosons modes.

\vspace{1cm}

\noindent
\textbf{\Large{Acknowledgments}}\\

\noindent
The authors would like to thank the referees for their constructive
comments. 
It is a pleasure to acknowledge discussions with J. Berges, D. Panja,
A. Patkos, J. Smit and A. Tranberg.
F.\,F. would like to thank P.\,J.\,H. Denteneer and D.\,F. Litim for
calling his attention to references \cite{glass} and
\cite{Baier:2000sb} respectively. The work of F.\,F. was supported by
FOM in the first part of this project,
N.\,D.\,A. by PPARC and P.\,S. by the Magnus
Ehrnrooth Foundation. 
This work is supported by the ESF Programme COSLAB - Laboratory
Cosmology and the NWO under the VICI Programme.

\end{document}